\documentclass{mem}
\pdfoutput=1

\usepackage{natbib}
\usepackage{txfonts}
\usepackage{balance}
\usepackage{graphicx}
\begin{document}
\def\teff{$T\rm_{eff }$}
\def\kms{$\mathrm {km s}^{-1}$}

\title{
Proper Motions in Globular Clusters using Deconvolution of HST Images
}

   \subtitle{}

\author{
Navtej\,Singh\inst{1}, 
Lisa-Marie\,Browne\inst{1}
\and Ray\,Butler\inst{1}
          }

  \offprints{Navtej Singh}

\institute{
Centre for Astronomy, National University of Ireland, Galway, University Road, Galway, Ireland.   \email{n.saini1@nuigalway.ie}
}

\authorrunning{Singh et al. }

\titlerunning{Proper Motions in Globular Clusters}

\abstract{
A sub-sampled deconvolution technique for crowded field photometry with the HST WFPC2 instrument was proposed by \citet{butler_2000} and applied to search for optical counterparts to pulsars in globular clusters \citep{golden_2001}. Simulations showed that the method, which takes account of the point-spread function (PSF) spatial variation, can provide better star detection and recovers the resolution lost to aberrations and poor sampling. The original emphasis was on precision photometry in crowded fields. In the present work, we have extended this technique to determine proper motions of stars in globular clusters. Actual HST images of the globular clusters, along with realistic simulations of the globular clusters, were used to benchmark the technique for astrometric accuracy. The ultimate aim is to use this dynamical data to search for intermediate-mass black holes in globular cluster cores.
}
\maketitle{}

%------------------------------------------------------------------------------------------------------------------------------------------------------------------------------
\section{Introduction}
Accurate detection and centroiding of stars are important for measuring stellar proper motions, in the search for intermediate-mass black holes (IMBH) in globular cluster cores \citep{gebhardt_2002,anderson_2010}. Observations and analysis of proper motions for many clusters are currently in progress \citep{lutzgendorf_2012b}. IMBH limits from HST proper motions have been presented for M15 \citep{McNamara_2003,vanderbosch_2006}, 47 Tuc \citep{mclaughlin_2006}, and Omega Centauri \citep{anderson_2010}. The high resolution and temporally consistent point spread function (PSF), over a suitably wide field, makes the HST a natural choice for this work. However the PSF of the HST WFPC2 and to a lesser extent ACS WFC cameras are under-sampled and strongly spatially varying. This reduces the statistical accuracy of the fit and causes errors in star detection. To recover the resolution lost to aberrations and poor sampling, \citet{butler_2000} developed the sub-sampled deconvolution technique for HST images.

%------------------------------------------------------------------------------------------------------------------------------------------------------------------------------
\section{Image Deconvolution}

\begin{figure*}[ht]
\centering
		\includegraphics[width=0.78\linewidth]{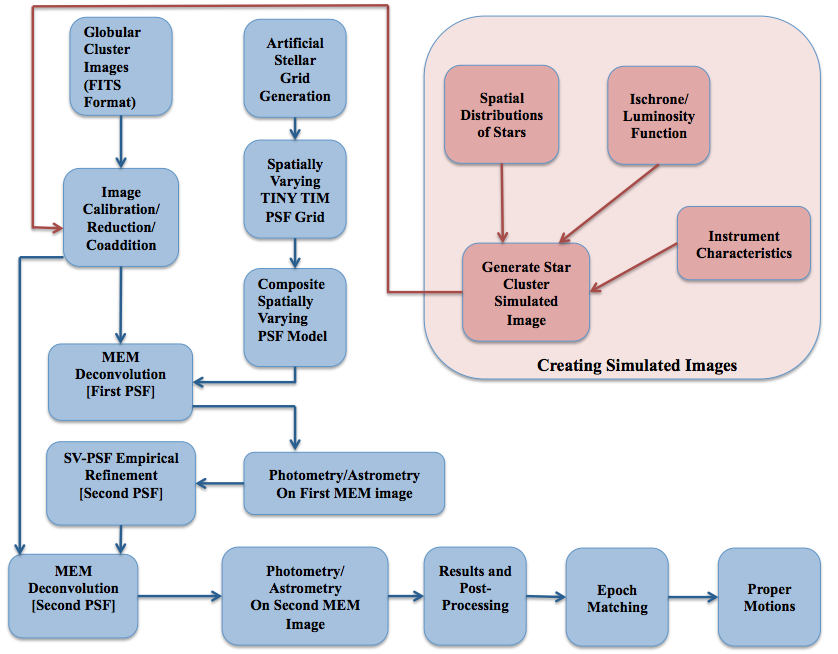}
		\caption{\footnotesize
				Stellar Proper Motion in Globular Clusters. Sub-sampled deconvolution recovers resolution and results in better star detection. The photometric and astrometric accuracy of the technique is benchmarked using realistic simulated globular cluster images.
		}
		\label{flow_chart}
\end{figure*}

\subsection{Sub-sampled deconvolution technique} 
We have developed an iterative method to extract an accurate, spatially varying point spread function (PSF) to be used for Maximum Entropy Method (MEM) based deconvolution. The flowchart in Fig. \ref{flow_chart}, outlines this sub-sampled deconvolution technique. An initial guess for the PSF is modeled from a sub-sampled spatially-varying TinyTim\footnote{Tool to model analytical HST PSFs. Refer to http://www.stsci.edu/hst/observatory/focus/TinyTim for more details.} PSF grid. Highly overlapping 256x256 pixel subimages of the science frame are deconvolved with the PSF appropriate for that position on the chip. The resulting deconvolved image is used for accurate star detection and first-pass photometry. 

Empirical improvements over the Tiny Tim PSF have been achieved by generating a second subsampled, and quadratically spatially varying PSF, modelled from the cluster image itself.  We use this in a second run through MEM.

Realistic simulations of WFPC2 cluster images, generated using Padova isochrones\footnote{Refer to http://stev.oapd.inaf.it/cgi-bin/cmd for more details.} and surface brightness profiles \citep{bertelli_2008}, have been used to benchmark the photometric and astrometric accuracy of the technique.

\subsection{Parallel Sub-Sampled Deconvolution} 
Sub-sampled deconvolution can be easily broken down into components to be run in parallel. Using native multiprocessing in Python, we have written a parallelized version of the sub-sampled image deconvolution algorithm \citep{singh_2012} that can be executed on multicore machines. Another version, using Message Passing Interface (MPI)\footnote{Software protocol used for parallel processing on distributed memory machines.} can be used on a cluster of machines. For benchmarking the computational speedup, we used a co-added HST WFPC2 PC1 chip image of the globular cluster NGC 6293, and a spatially varying analytical PSF model was generated from the Tiny Tim PSF grid. A  large speedup was found when a high fraction of code is running in parallel.

\subsection{Star Detection and Photometry}
Stars were detected, and measured by aperture photometry, on the deconvolved images as well as on the original, reduced and combined images. Depending on the degree of crowding, IRAF\footnote{IRAF is distributed by the National Optical Astronomy Observatories,
    which are operated by the Association of Universities for Research
    in Astronomy, Inc., under cooperative agreement with the National
    Science Foundation.} DAOFIND and SExtractor \citep{bertin_1996} were used. Star positions were corrected for the 34th row defect \citep{anderson_1999} and the optical distortions \citep{anderson_2003}.

\subsection{Proper Motion}
Milli-arcsecond absolute proper motion is not possible with HST WFPC2 \citep{mclaughlin_2006}. Relative stellar proper motions were determined by transforming star positions from each epoch to a reference-frame epoch. The {\sc IRAF} task \textit{geomap} was used to transform the coordinates using a second order polynomial solution, which takes care of the residual geometric distortions.

%------------------------------------------------------------------------------------------------------------------------------------------------------------------------------
\section{Results}
\begin{figure}[h]
\centering
		\includegraphics[width=0.45\linewidth]{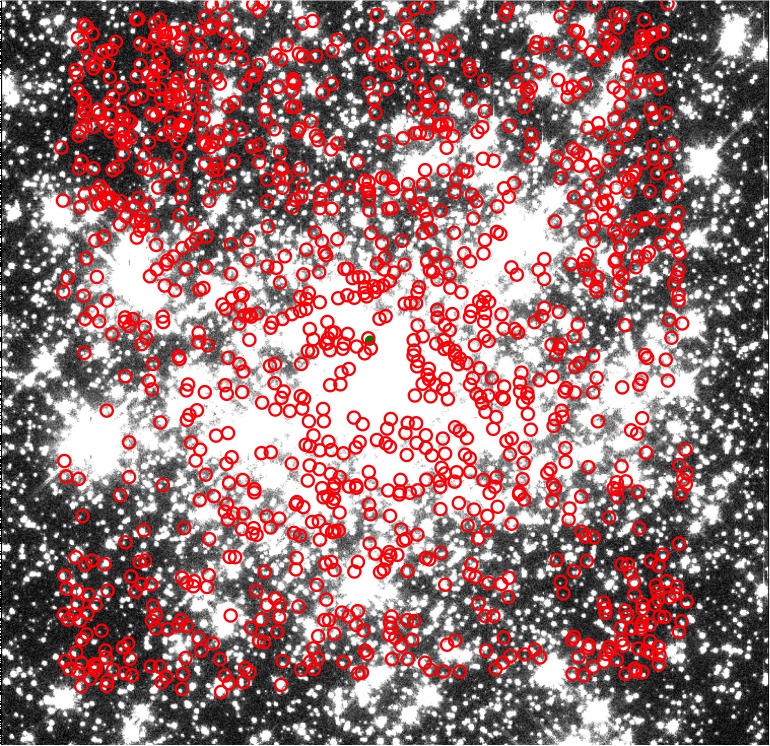}
		\includegraphics[width=0.45\linewidth]{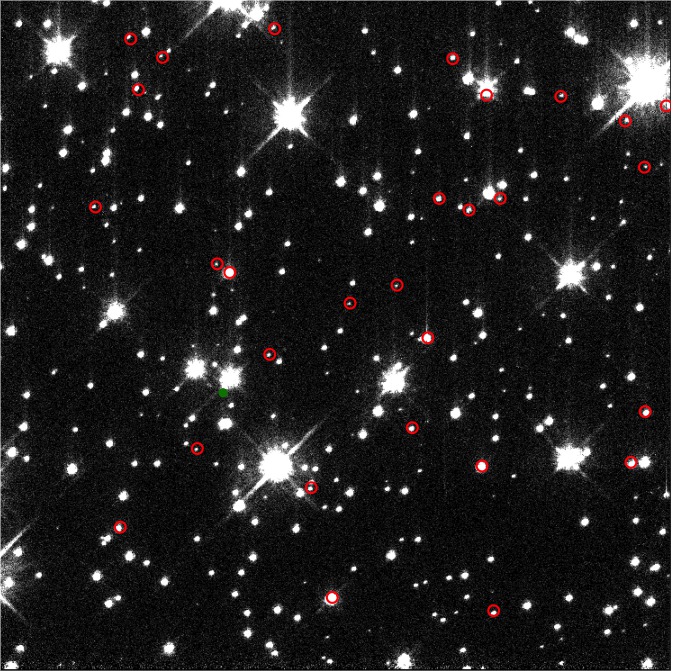}

		\caption{
				\footnotesize
				NGC 6293 on the left and M71 on the right. The circles mark the stars with the largest motions between epochs. The clustering of such stars towards the corners of the NGC 6293 images indicates residual systematics in the distortion mapping.
		}
		\label{cluster_images}
\end{figure}

\begin{figure}[!h]
\centering
		\includegraphics[width=0.85\linewidth]{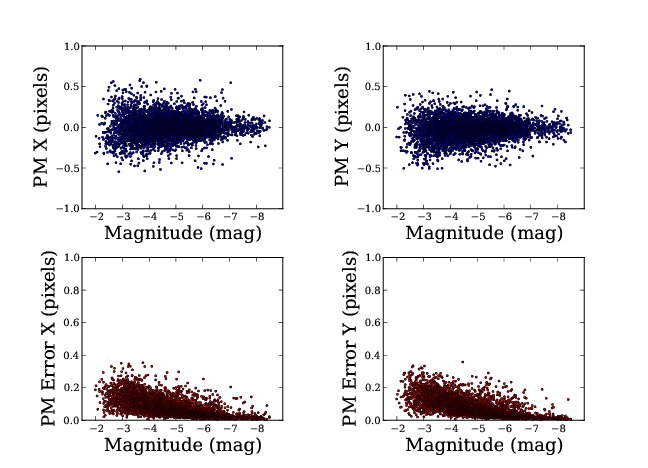}
		\includegraphics[width=0.85\linewidth]{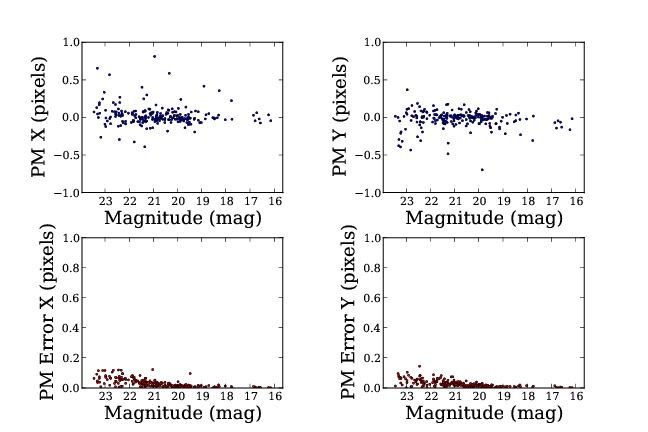}
		\caption{
		\footnotesize
			NGC 6293 on the top and M71 on the bottom. X and Y proper motions in units of pixels (top) and error estimates (bottom). Errors are derived from the consistency of results obtained by processing several individual frames in each epoch, as per \citet{anderson_2006}. 
		}
		\label{pm_mag}
\end{figure}

\begin{figure}
\centering
		\includegraphics[width=0.85\linewidth]{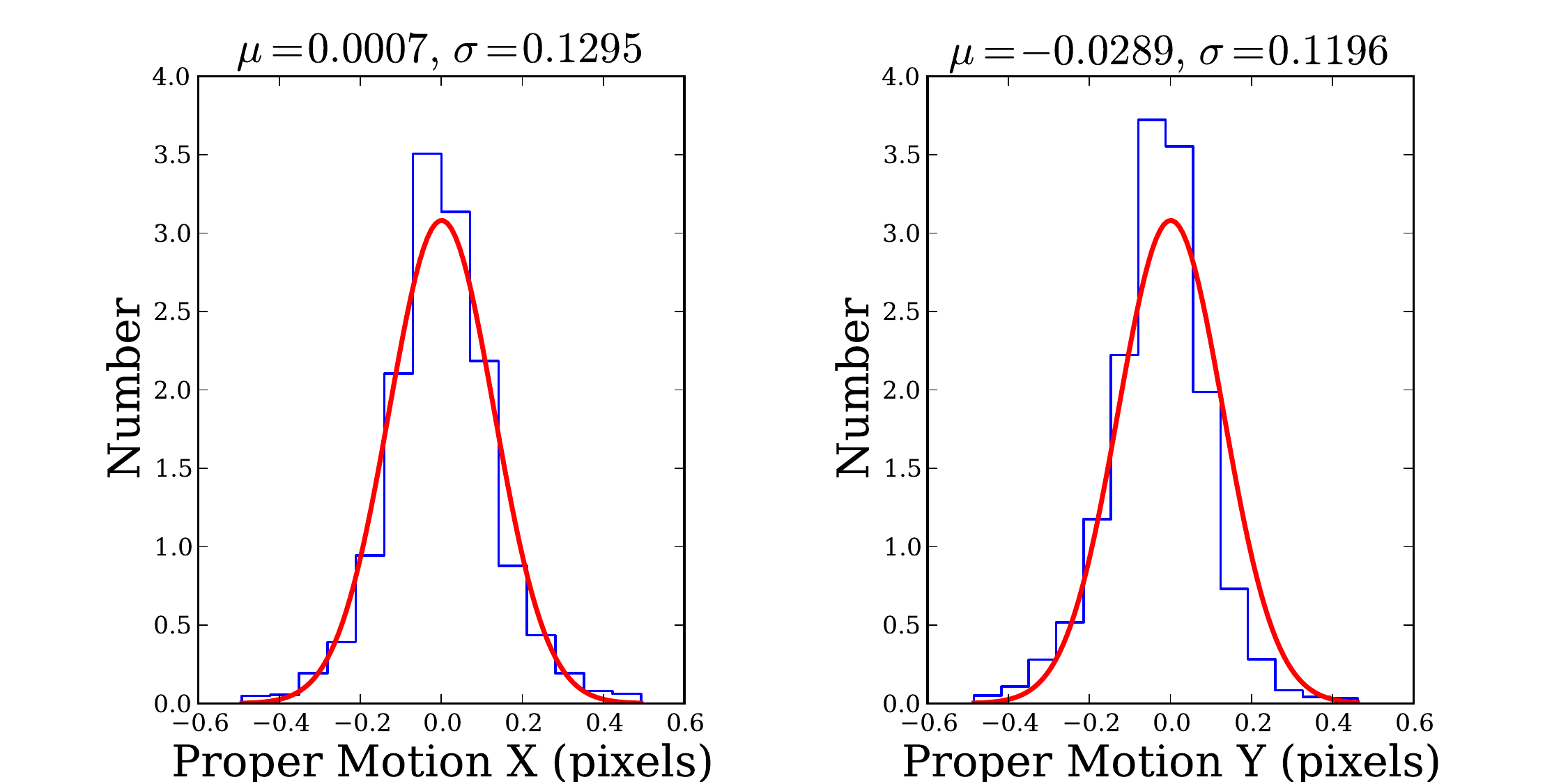}
		\includegraphics[width=0.85\linewidth]{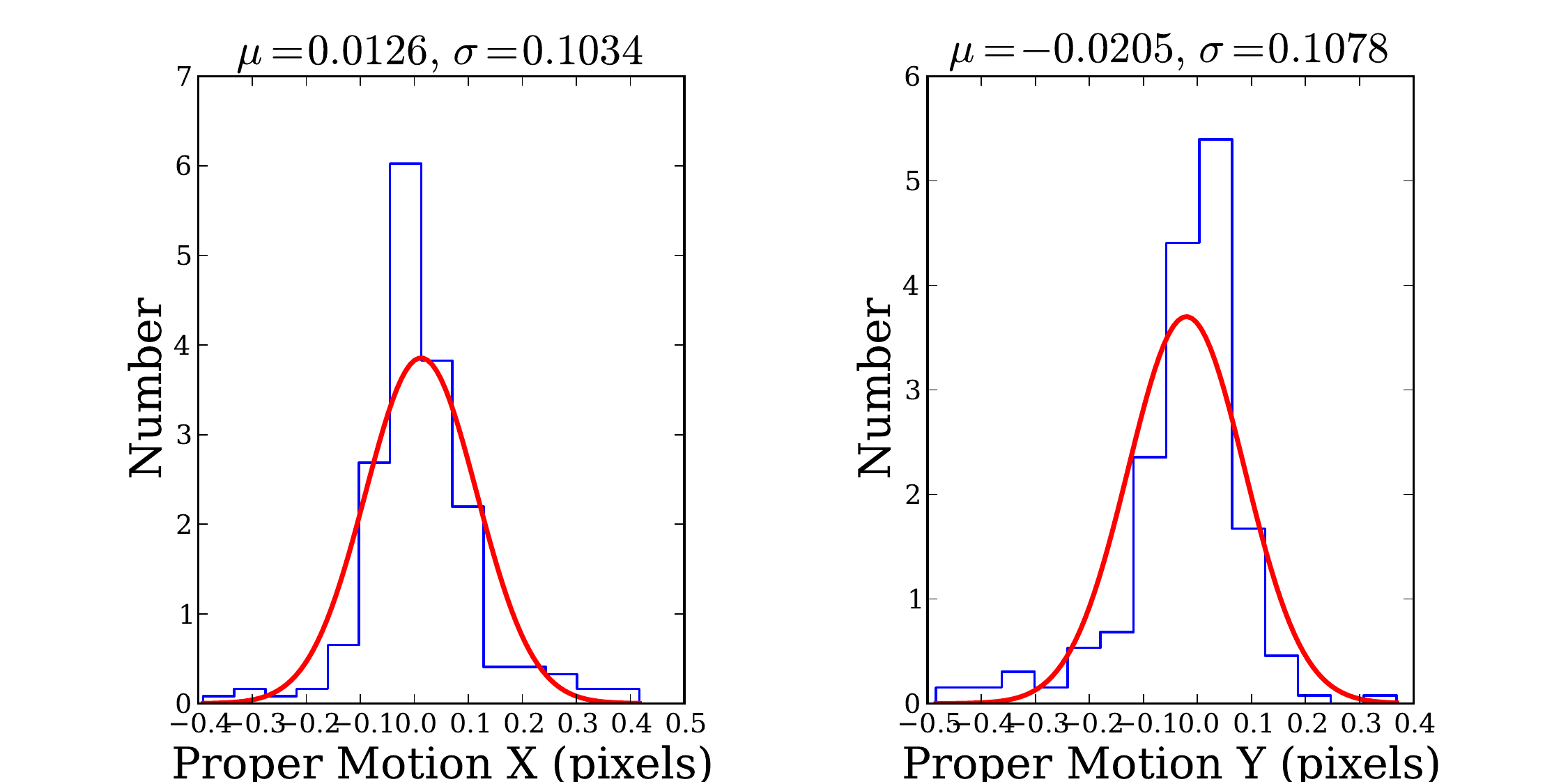}
		\caption{
		\footnotesize
			NGC 6293 on the top and M71 on the bottom. Histograms of X and Y proper motion in units of pixels, with Gaussian fits.
		}
	\label{pm_hist}
\end{figure}

\begin{figure}
\centering
	\includegraphics[width=0.76\linewidth]{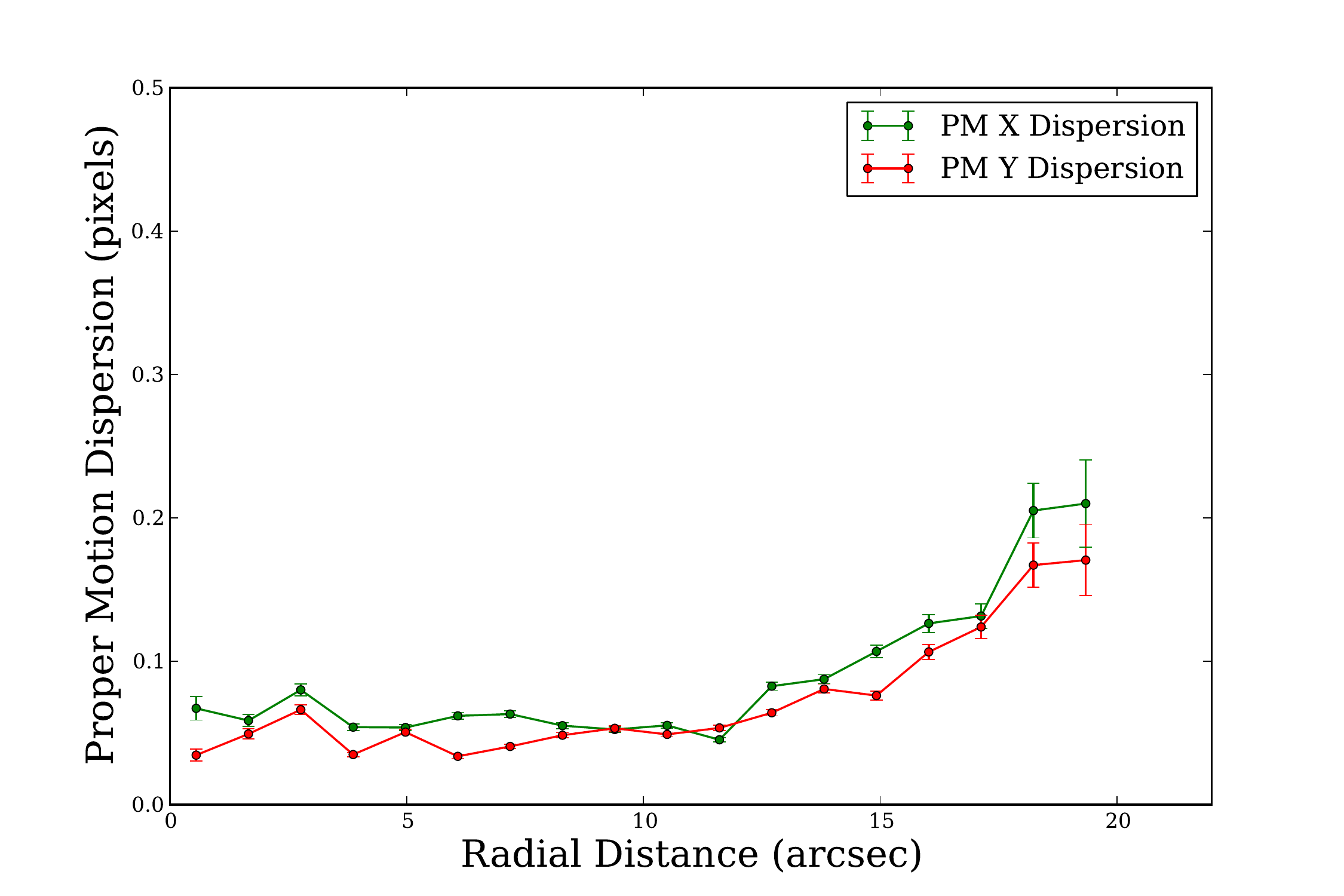}
		\includegraphics[width=0.76\linewidth]{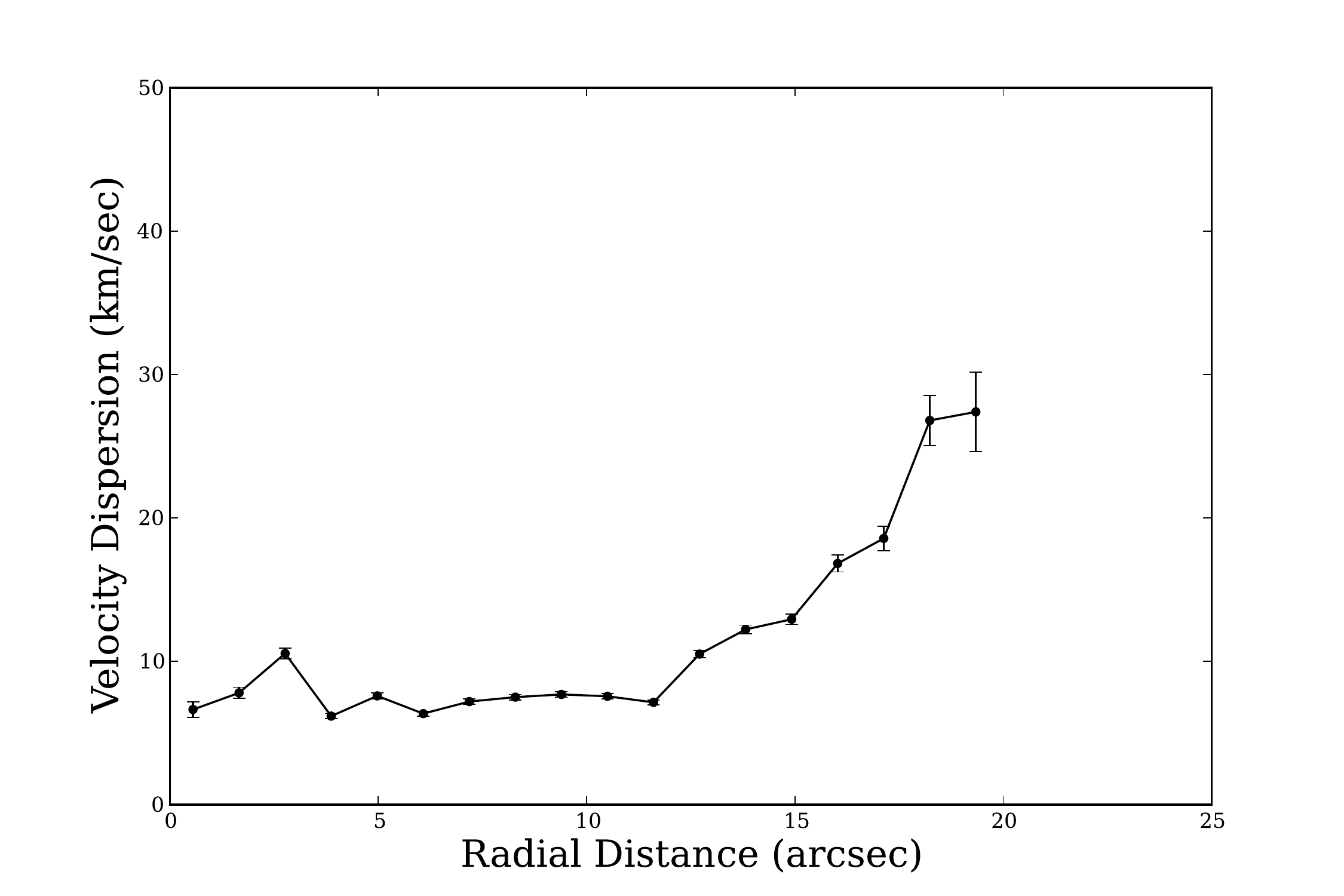}
		\caption{
		\footnotesize
		NGC 6293 only. Proper motions, binned in steps of 1 arcsec, versus distance from cluster centre. Top: X and Y motions in units of pixels. Bottom: Total motion in units of velocity. The increased values at radii $>$ 12 arcsec reflect the corners issue shown above.
		}
		\label{pm_radial}
\end{figure}

We show preliminary results for 2 epochs of NGC 6293 and M71 in the F555W filter of the WPFC2 camera on the HST.  Our  results for M71 agree closely with \citet{samra_2012}. The high velocity stars are marked for both the clusters in Fig. \ref{cluster_images}.  Proper motion as a function of magnitude is shown in  Fig. \ref{pm_mag}. A gaussian fit to the proper motion histogram is shown in Fig. \ref{pm_hist}. Velocity dispersion as a function of radial distance from the center of the cluster is shown in Fig. \ref{pm_radial}.

%------------------------------------------------------------------------------------------------------------------------------------------------------------------------------
\section{Conclusions}

%Future Work
In these preliminary results, we found no signature of a central black hole in the globular clusters M71 and NGC 6293. We are now extending the proper motion study to deconvolved cluster images taken with the Advanced Camera for Surveys (ACS) camera on the HST. The spatially varying, sub-sampled deconvolution technique is also being adapted for adaptive optics images. Wide-field adaptive optics, with large telescopes, open a new frontier in determining accurate parameters for most globular clusters that remain essentially unstudied because of high reddening, crowding, and large distances \citep{Ortolani_2011}.

%------------------------------------------------------------------------------------------------------------------------------------------------------------------------------d
\begin{acknowledgements}
Based on observations made with the NASA/ESA Hubble Space Telescope in programs GO-5899, GO-10775, and GO-11975, obtained from the data archive at the Space Telescope Institute. STScI is operated by the association of Universities for Research in Astronomy, Inc. under the NASA contract NAS 5-26555. We gratefully acknowledge the support of Science Foundation Ireland (under award 08/RFP/PHY1236) and the College of Science, NUI Galway (under their PhD fellowship scheme).
\end{acknowledgements}

%------------------------------------------------------------------------------------------------------------------------------------------------------------------------------
\bibliographystyle{aa}

\end{document}